\documentclass{aa}
\usepackage{graphicx}
\begin{document}
   \title{The Degree of CO Depletion in Pre-stellar Cores
\thanks{Based on observations collected at the IRAM 30\,m, Spain}}

\author{A. Bacmann
          \inst{1}\thanks{present address: abacmann@eso.org},
          B. Lefloch\inst{2}
          C. Ceccarelli\inst{3,2},
          A. Castets\inst{3},
          J. Steinacker\inst{1},
          L. Loinard\inst{4}
          }

\offprints{A. Bacmann}

\institute{Astrophysikalisches Institut und Universit\"ats Sternwarte 
(AIU), Schillerg\"a{\ss}chen 2-3, D-07745 Jena, Germany\\
\email{bacmann@astro.uni-jena.de, stein@astro.uni-jena.de}
\and
Laboratoire d'Astrophysique, Observatoire de Grenoble, BP\,53, F-38041 
Grenoble C\'edex 9, France\\
\email{lefloch@obs.ujf-grenoble.fr}
     \and
Observatoire de Bordeaux, 2 rue de l'Observatoire, BP\,89, F-33270 Floirac, 
France\\
\email{ceccarel@observ.u-bordeaux.fr, castets@observ.u-bordeaux.fr}
     \and 
Instituto de Astronom\'{\i}a, Universidad Nacional Aut\'onoma de M\'exico, 
Apartado Postal 3-72 (Xangari), 58089 Morelia, Michoac\'an, Mexico\\
\email{l.loinard@astrosmo.unam.mx}
}
\date{Received 5 March 2002 / Accepted 21 March 2002}

\abstract{We present new results on CO depletion in a sample of nearby 
pre-stellar cores, based on observations of the millimeter 
C$^{17}$O and C$^{18}$O lines and the 1.3\,mm dust emission with the IRAM 30\,m
telescope. In most cases, the distribution of CO is much flatter 
than that of the dust, whereas other tracers, like $\rm N_2H^{+}$, 
still probe the latter. In the centre of these objects, we estimate CO to 
be underabundant by a factor 4-15 depending on the cores. The CO
underabundance is more pronounced in the central regions and appears to 
decrease with increasing distance from the core centre. This
underabundance is most likely due to the freezing out of CO onto the dust 
grains in the cold, dense parts of the cores. 
We find evidence for an increase of the CO depletion degree with the 
core density.
}
\authorrunning{Bacmann et al.}
\maketitle
\def\ceio{\hbox{C$^{18}$O}}
\def\c17o{\hbox{C$^{17}$O}}
\def\htwo{\hbox{$\rm H_{2}$}}
\def\cmmt{\hbox{\kern 0.20em cm$^{-3}$}}
\def\cmpd{\hbox{\kern 0.20em cm$^{2}$}}
\def\kms{\hbox{\kern 0.20em km\kern 0.20em s$^{-1}$}}
\def\gmu{\hbox{\kern 0.20em g$^{-1}$}}


\section{Introduction}

Of all the molecules present in the interstellar medium, CO (and its isotopes)
 has been the object of particular attention. It has been widely used to 
study the mass 
distribution and kinematics of star forming molecular clouds. Moreover, CO 
is thought to play a major role in the thermal cooling of molecular clouds 
(e.g., Goldsmith \& Langer, \cite{goldsmithlanger}). In very dense and cold 
regions of the 
interstellar medium the CO molecules tend to leave 
the gas phase and condense out onto the dust grains, directly affecting 
the thermal balance of clouds (Goldsmith, \cite{goldsmith}). CO also 
participates in a complex chemical network and its depletion leads to
changes in gas composition and chemical reactions. These, in turn, can 
influence  the contraction of molecular clouds 
through ambipolar diffusion (e.g. 
Shu et al. \cite{shu87}), via the ionisation fraction 
(Caselli et al. \cite{caselli98}) or the gas phase deuteration of molecules
(Roberts \& Millar \cite{roberts}a, \cite{roberts}b). 

This phenomenon affects in particular pre-stellar condensations, which are
characterized by very low temperatures ($\sim$ 10\,K) and high central
densities ($\sim$ $10^4-10^5$\,cm$^{-3}$). These objects, still starless, 
are thought to evolve slowly
via ambipolar diffusion towards higher degrees of condensation until they
become unstable and collapse under their own gravity to form a protostar.
This pre-collapse phase is believed to strongly influence the following stages
of the star formation process - mass infall rate (e.g. Henriksen et al. \cite{hab}), 
fragmentation, mass of future star (e.g. Bacmann et al. \cite{bacmann}).
It is therefore of prime importance to characterize the physical and chemical
conditions in the gas and dust of pre-stellar condensations which have not yet 
started to collapse. The study of CO depletion is in this context particularly relevant 
due to the implications of e.g. modified thermal balance or contraction by ambipolar
diffusion induced by a decreased abundance of the molecule.
CO depletion in dense cores has already been discussed by Kramer et al. (\cite{kramer}), 
Willacy et al. (\cite{willacy}), and Caselli et al. (\cite{caselli}, \cite{cas02a}, \cite{cas02b}) who 
focussed on L1544, a core similar to the ones we discuss in this work. Here,
we studied the distribution and abundance of CO via its rarer isotopes 
C$^{17}$O(1-0) and C$^{18}$O(1-0), and of N$_2$H$^+$(1-0) in a sample of 
pre-stellar cores, using the 1.3\,mm continuum as a tracer of the H$_2$ 
column density. Our aim was to quantify the depletion and to link it with 
physical properties of the cores. We first briefly present the observations
in Section\,\ref{obs}, then the spatial distributions and column density 
estimations in Section\,\ref{res}, discuss our results and conclude in
Section\,\ref{disc}.


\section{Observations and data reduction}

\label{obs}

We have selected a sample of 7 pre-stellar cores previously studied 
by Bacmann et al. (\cite{bacmann}, cf. Table\,\ref{sources}). The core
L1544, studied in detail by 
Caselli et al. (\cite{caselli}, \cite{cas02a}, \cite{cas02b}), has also been included for comparison.

\begin{table*}[!]
\caption{Comparison of C$^{17}$O and  (calculated from dust) H$_{2}$
properties. The \htwo\ densities are taken from Table 3 of Bacmann et al. 
(2000). Errors
on the temperature are typically $\pm 1$\,K. Uncertainties on $N_{\rm H_2}$ and
$n({\rm H_{2}})$ are a factor of $\sim 2$, those on $X$ and $f$ a factor of
 $\sim 2.5$.}
\begin{tabular}{lccccccccc}
\hline
Name     & $\int T_{\rm mb}^{10}({\rm C^{17}O}) dv$ & $T\rm_k$ & 
$N_{{\rm C^{17}O}}$ & $N_{\rm H_{2}}$ & $n(\htwo)$ & 
$N_{{\rm C^{17}O}}/N_{\rm H_{2}}$ & $f$ & $f^{70}$ & $f^{50}$\\
       & $\rm K\kms$   &  (K) & (cm$^{-2}$)    & (cm$^{-2}$) & ($\cmmt$) & $ (=X)$ &
 $(=X_{\rm can}/X)$   & & \\
\hline
L1544  &  $-$             & 10  & 5.6$\pm 0.8\times 10^{14}$ & 1.6$\times 10^{23}$  & 4$\times 10^5$   & 3.5$\times 10^{-9}$ & 14   & $-$ & $-$\\
L1689B & 1.42($\pm 0.14$) & 9.3 & 1.5$\pm 0.2\times 10^{15}$ & 1.4$\times 10^{23}$  & 1.4$\times 10^5$ & 1.1$\times 10^{-8}$ & 4.5  & 2   & 1.5\\
L1709A & 1.18($\pm 0.12$) & 10  & 1.2$\pm 0.2\times 10^{15}$ & 1.4$\times 10^{23}$  & 1.2$\times 10^5$ & 8.6$\times 10^{-9}$ & 5.5  & 2.5 & 2\\
L310   & 0.35($\pm 0.03$) & 8.8 & 2.8$\pm 0.5\times 10^{14}$ & 7.3$\times 10^{22}$  & 9$\times 10 ^4$   & 3.8$\times 10^{-9}$ & 12.5 & 5   & 3\\
L328   & 0.60($\pm 0.06$) & 9.7 & 5.3$\pm 0.8\times 10^{14}$ & 9.5$\times 10^{22}$  & 1.8$\times 10^5$ & 5.6$\times 10^{-9}$ & 8.5  & 5   & $-$\\
L429   & 0.50($\pm 0.05$) & 7.5 & 4.3$\pm 0.8\times 10^{14}$ & 1.4$\times 10^{23}$  & 6$\times 10^5$   & 3.1$\times 10^{-9}$ & 15.5 & 5   & 3\\
Oph D  & 0.57($\pm 0.06$) & 8.2 & 4.8$\pm 0.9\times 10^{14}$ & 1.4$\times 10^{23}$  & 3$\times 10^5$   & 3.4$\times 10^{-9}$ & 14   & 4.5 & 3\\
\hline
\end{tabular}
\label{sources}
\end{table*}

Observations in the millimeter lines of 
N$_2$H$^+$(1-0), C$^{17}$O(1-0), $\ceio$(2-1) and (1-0) were 
made in August 2000 and 
2001 at the IRAM 30\,m telescope. The spectra were taken in position 
switching mode with a 20\arcsec\ sampling along cuts in chosen directions
passing through the maximum of the 1.3\,mm 
emission. 
The reference positions were checked to 
be free of C$^{18}$O(1-0) emission. We used an autocorrelator with a  
resolution of 20\,kHz as backend. The spectral resolution was degraded 
so as to provide a kinematic resolution of $\sim 0.1\kms$ at all wavelengths. 
Pointing was checked typically every 1.5 hour and found to be 
better than $\sim 3\arcsec$. The data were reduced 
using the CLASS package.
The line fluxes are expressed in units of main-beam brightness temperatures.  
The uncertainties in the fluxes are $\approx 10\%$. 
The beam size ($\sim 22\arcsec$ at 110\,GHz) and the efficiencies (FWHM) 
are taken from the IRAM web page http://www.iram.es/. Previous mapping of
the cores in the 1.3 millimeter continuum is described in Bacmann et al. 
(\cite{bacmann}).

%

\section{Results}

\label{res}

\subsection{Distribution of molecular and dust tracers}

\begin{figure*}[!ht]
\centering
\vspace{-3cm}
\includegraphics[width=17cm]{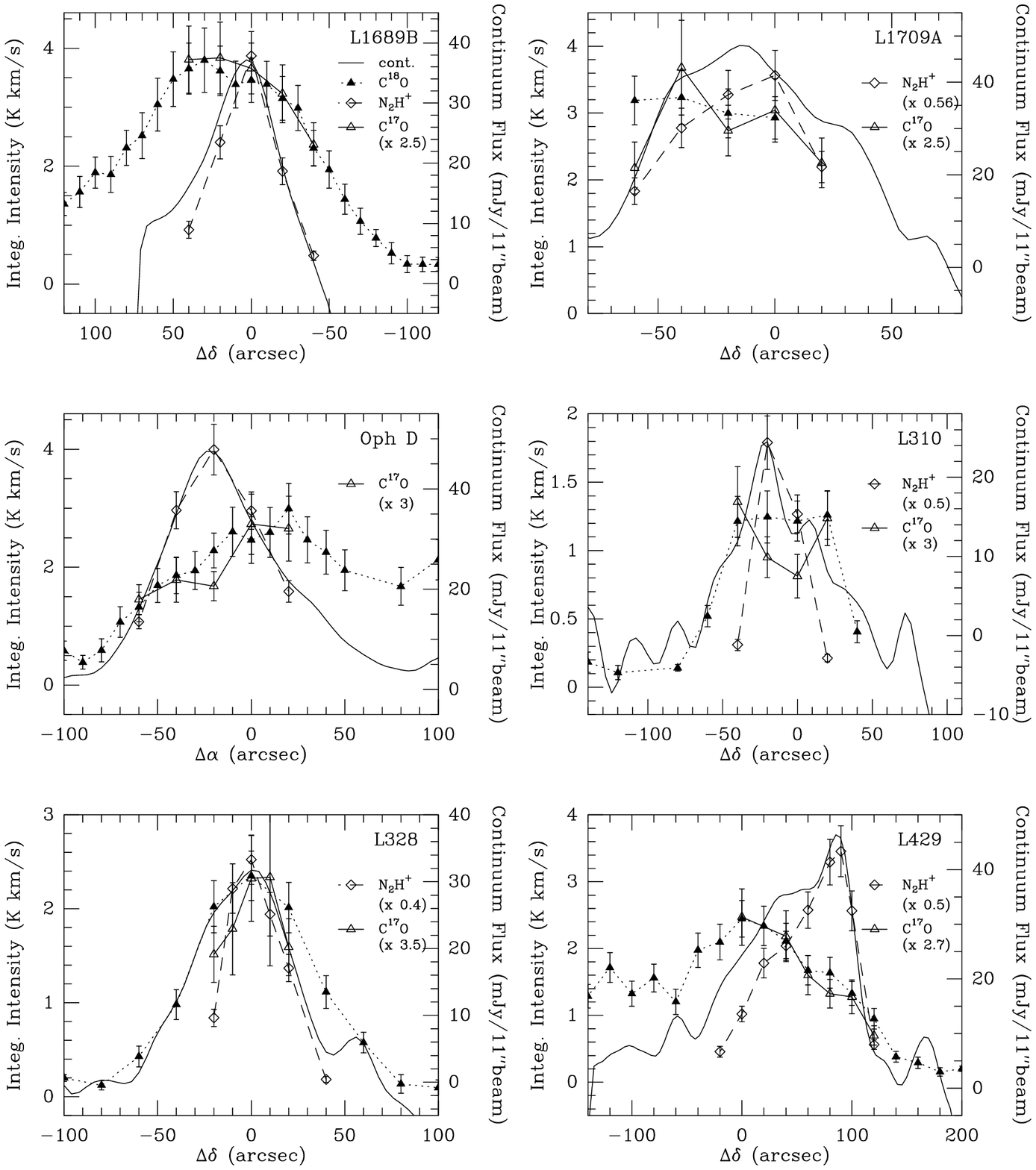}
\vspace{-2.5cm}
\caption{Spatial distribution of the integrated line intensity of 
C$^{18}$O(1-0) (filled triangles), C$^{17}$O(1-0) (empty triangles), 
N$_2$H$^+$(1-0) (empty diamonds), and of the dust continuum emission (solid
line) at 1.3\,mm, along cuts in R.A., Dec, or the diagonal 
$\Delta\alpha=\Delta\delta$ (for L310 and L1709A) passing through the 
maximum of the continuum emission.}
\label{cut}
\end{figure*}

We compared the spatial distribution of CO, N$_2$H$^+$ and of the dust traced 
by the 1.3\,mm continuum emission. To this effect, we plotted for each core 
the integrated intensity of the molecular lines (C$^{18}$O(1-0), 
C$^{17}$O(1-0) and N$_2$H$^+$(1-0)) together with the millimeter continuum 
flux, along directions passing through (or close to) the maximum of the 
1.3\,mm continuum map. To obtain the integrated line intensities, we first 
fitted a gaussian (or a series of gaussians in the case of hyperfine 
structures) to the considered line, and then evaluated the area under the 
gaussian. The results obtained are similar if we consider the integrated 
signal between two given velocities. The resolution of the 1.3\,mm continuum 
map was degraded to the resolution of the molecular line observations, 
22\arcsec. The resulting plots are presented in Fig.\,\ref{cut} for all cores 
except L1544. Whereas the continuum is sharply peaked (except for L1709A), the 
distribution of the C$^{18}$O(1-0) and C$^{17}$O(1-0) integrated intensities is
flatter (e.g. L1689B and L310) or completely fails to trace 
the strong continuum 
peak (e.g. L429 and Oph\,D). The core L328 is particular in the sense that the 
CO isotopes are also sharply peaked (their distribution is only slightly 
broader than that of the dust: $\sim$\,15\% broader at half-maximum). For all 
cores, the distribution of N$_2$H$^+$, on the other hand, closely follows that 
of the dust traced by the continuum.

These results tend to show that CO is underabundant with respects to H$_2$ 
(traced by the continuum) in the centres of the cores. As Caselli et al. 
(\cite{caselli}) pointed out, this is most likely due to the depletion of 
CO onto the dust grains, because of the high densities and low temperatures 
in the inner parts of the cores. However, it can already be seen from the 
various behaviours of the plots such as those in Fig.\,\ref{cut} 
that the amplitude of this phenomenon, the depletion degree, 
is not the same in all the cores considered.

\subsection{Physical Conditions}

The average \htwo\ density at the emission peak 
in the cores was estimated to be of a few $10^5\cmmt$ in the 11\arcsec\
telescope beam by 
Bacmann et al. (2000, see also Table\,\ref{sources}). Hence, the populations of the 
energy levels of CO and its isotopes are very close to thermalization.
The ratio of the \c17o\ (2-1) to (1-0) line integrated areas, 
calculated under the assumptions of optically thin lines and at LTE, 
therefore provides a good estimate of the kinetic 
temperature. We applied the same procedure to the \ceio\ lines under the 
same assumptions. We obtained consistent results between both isotopes, 
with values in the range 6-8\,K for the \c17o\ lines and 
in the range 7.5-10\,K for \ceio. The results are summarized in Table\,\ref{sources}.
Such temperatures are  similar to those 
found towards other pre-stellar and/or protostellar cores 
(Caselli et al. 1999; Fuller \& Myers, 1993; Ladd, Fuller \& Deane, 1998). 

We evaluated the CO column density using the rare isotope C$^{17}$O
in order to make sure our results were not affected by opacity effects. 
The opacity of the C$^{18}$O(1-0) transition was determined following the 
method described by Ladd et al. (\cite{ladd}). 
Supposing a [$^{17}$O]/[$^{18}$O] abundance 
ratio of 3.65 (Wilson \& Rood \cite{wilson}; Penzias \cite{penzias}), we 
derived opacities lower than 0.1 for Oph\,D, L328 and L310, of 1.1 for L429, 
1.3 and 1.5 for L1709A and L1689B, respectively.  According to the 
aforementioned C$^{18}$O(1-0) opacities, the C$^{17}$O(1-0) transition is 
optically thin for all of the cores. This validates our determination of the 
kinetic temperature from the \c17o\ lines.
Although 
the opacity effects are small in the general case, we prefered to use 
the same tracer C$^{17}$O for {\em all} cores for more accurate comparisons. 
The C$^{17}$O column density was determined under the same hypothesis and
adopting as kinetic temperature the values derived previously. 
The values of $N\rm _{C^{17}O}$ determined at the continuum emission peak
are presented in Table\,\ref{sources}. 

The H$_2$ column density was determined from the dust thermal 
emission at 1.3\,mm, which is largely optically thin.
We degraded the resolution of all our continuum maps to the resolution at
the frequency of the $\rm C^{17}O$(1-0) line, so as to compare H$_2$ 
and C$^{17}$O column 
densities in the same beam. 
The dust temperature was taken to be 7.5\,K in all cores at their centre,
as suggested by the recent works of Zucconi et al. (\cite{zucc}) and 
Evans et al. (\cite{evans}). This is slightly lower than in the envelope, 
where values of 11\,K were determined from ISOPHOT observations 
(Ward-Thompson et al., \cite{ward01}).
Following Andr\'e et al. (1996), we have adopted  a dust opacity
coefficient $\kappa_{1.3\rm mm}= 0.005\cmpd\gmu$, appropriate for gas in the 
central pre-stellar regions, where the \htwo\ density is $\sim 10^5\cmmt$, 
though the actual value is uncertain by a factor of 2 (Preibisch et al., 
\cite{preibisch}).  
The values of the H$_2$ 
column density at the maximum of the 1.3\,mm continuum emission are shown in 
Table\,\ref{sources}. 

\section{Discussion}

\label{disc}


The results presented above show strong evidence for CO depletion onto the 
dust grains in the densest parts of the molecular cloud cores. 
We present in Table\,\ref{sources} the ratio $X$ of the C$^{17}$O and the 
H$_2$ column densities. For all cores, the abundance $X$ 
found is lower than the ``canonical'' abundance determined by Frerking et al. 
(\cite{frerking}) towards dark cores $X_{\rm can}= 4.8\times 10^{-8}$ 
and varies from 4.5 times smaller to nearly 16 times
 smaller than $X_{\rm can}$. 
The CO molecules thus appear underabundant with respect to 
\htwo\ at the centre of the pre-stellar cores. The values found suggest 
different 
depletion degrees for the different cores, as was already hinted at by the 
cuts of Fig.\,\ref{cut}. The depletion factor $f$, defined as 
$X_{\rm can}/X$, is 
also shown in Table 1. For L1544,  we derive a depletion factor of 14, 
the difference with the value of 10 found by Caselli et al. (\cite{caselli}) 
being due to our 
assumption of a colder central dust temperature. 
Adopting a higher value for the dust absorption coefficient $
\kappa_{1.3\rm mm}$, 
representative of denser environments would decrease the depletion factor 
to 6 in Oph\,D, L429, L310. This case is extremely unlikely as it requires
typical densities of $\sim 10^6\cmmt$ over a size of 20\arcsec\ (3200 AU). 
The
depletion factor would also be a factor of $\sim 2$ smaller if the central
temperature is taken to be 11\,K instead of 7.5\,K. 
These results considerably extend the previous works on depletion in
pre-stellar cores by Kramer et al. (1999), Caselli et al. (1999) and 
more recently by Tafalla et al. (2002). 
The latter report similar depletion
factors for L1544 and 4 other pre-stellar cores in the Taurus molecular
cloud.

The effect of CO underabundance is still detected at 
large scale, with a smaller amplitude, in regions covering large parts of 
the lower density gas, where the ``standard'' value of 
$\kappa_{1.3\rm mm}$ ($0.005\cmpd\gmu$) is a reasonable approximation. 
For each core, we derived the C$^{17}$O and H$_2$ column density along our 
cuts at 70\% and 50\% of the maximum of the continuum emission. 
The values of 70\% and 50\% were chosen 
as a  compromise between having regions of lower density and not being 
affected by the loss of sensitivity of the millimeter continuum at the edges 
of the 1.3\,mm maps (e.g. Andr\'e et al. \cite{awb}). We assumed higher dust 
temperatures of 10\,K and 11\,K to derive the depletion factors 
$f^{70}$ and $f^{50}$, respectively, since the outer layers are 
thought to 
be warmer than the core centre (e.g. Zucconi et al. \cite{zucc}).
The abundance $X^{70}$  is 2 to 5 times 
smaller than $X_{\rm can}$, {\em the CO underabundance is more pronounced 
in the densest, central regions}. For all cores, the average depletion 
factor $f^{70}$ is lower than $f$, and $f^{50}$ is lower than 
$f^{70}$. 

Such variations  can be easily understood if depletion 
is the cause of this underabundance of CO. Indeed, the less dense and 
warmer parts of the cores should be less affected by the freezing out of 
CO molecules onto dust grains. Within the objects, the variations in density 
are about one order of magnitude, based on the density fits of Tafalla et 
al. (2002), whereas the temperature varies by $\sim 50\%$ at most. Hence, 
the variations of the depletion degree are probably dominated by the 
variations in density and not in temperature. We stress the fact that the
relative variations between $f$, $f^{70}$ and $f^{50}$  remain true 
even if the temperature is assumed uniform in the cores.
This relation shows that the depletion factor increases with density inside of 
the core, independently of any modelling of the internal core structure.
The increase in $f$ is particularly pronounced in the case of the densest 
cores L429 and Oph\,D. 

\begin{figure}[ht]
\centering
\includegraphics[width=7cm,angle=0]{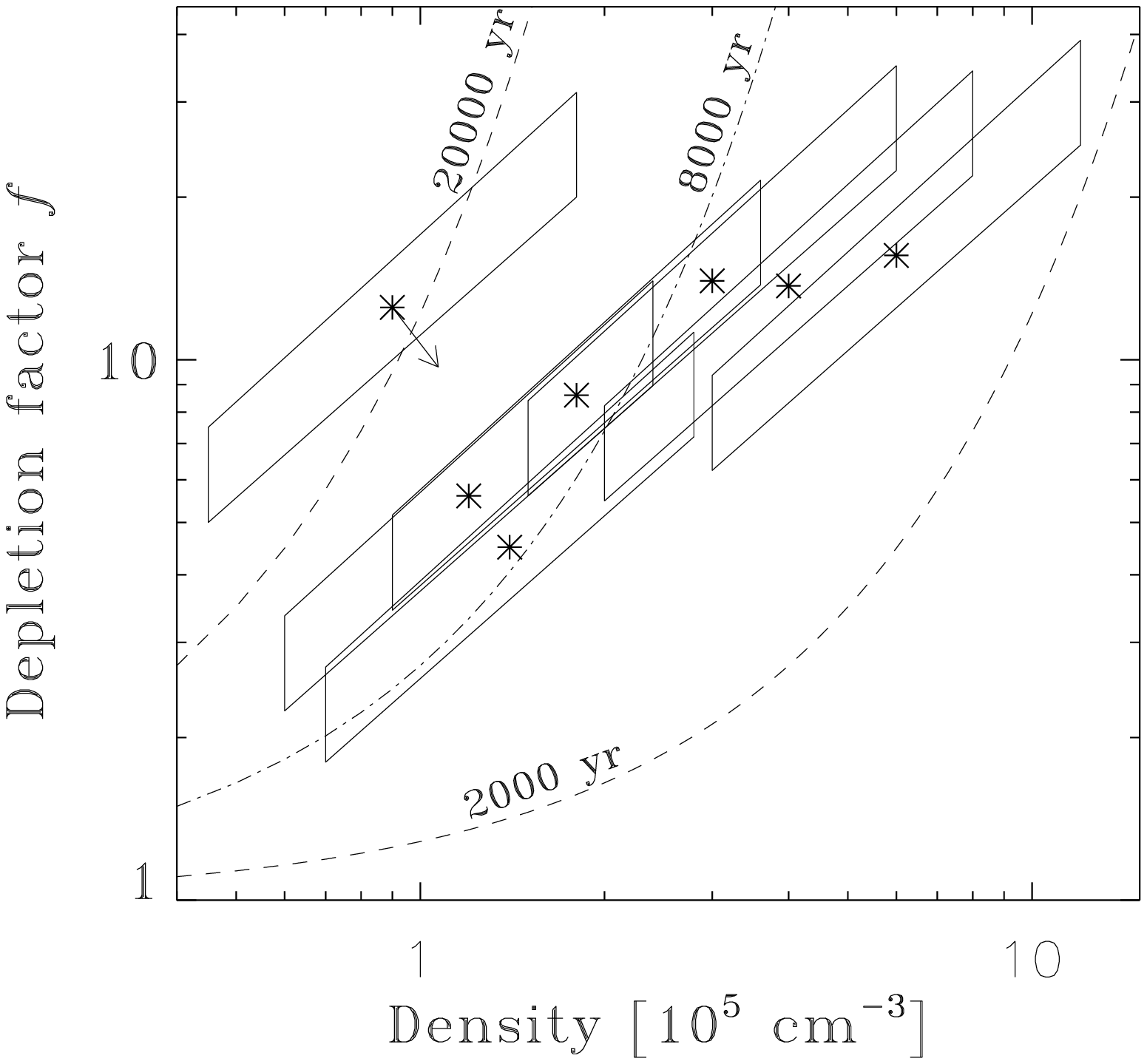}
\caption{Variations of the depletion factor $f= X_{\rm can}/X$
as a function of the mean central density. The parallellograms represent 
the error boxes for each point, accounting for the fact that the error on 
$\kappa_{1.3\rm mm}$ affects both density and depletion estimates 
simultaneously. The arrow indicates the direction the point for L310 would 
move with better density estimates and more reliable C$^{17}$O measurements.
The dashed and dash-dotted lines represent the simple model 
$f=\exp(t/\tau_{\rm d})$, where $\tau_{\rm d}$ is the depletion 
timescale, for various core ages.}
\label{density}
\end{figure}

We tried to relate the CO underabundance with the physical properties of the 
core. 
To this effect, we plotted the depletion factor $f$ as a function of the 
mean density in the core central region (Fig.\,\ref{density}). 
The above plot indicates an increase of the depletion factor with 
density, confirming the trend observed for individual cores.
A linear regression of the form $f\propto n(\htwo)^{\alpha}$ performed 
on the {\em whole} data set yields $\alpha\simeq 0.1-0.5$. The rather large
range of values allowed for the exponent is mainly due to the point L310 
which stands out rather markedly of the trend mentionned above. 
Indeed, leaving aside the L310 data, the points match rather well a relation 
of the form $f\propto n(\htwo)^{0.7-0.8}$. 
We note that the observational uncertainties are much larger on L310 
than on the rest of the dataset. First, the weather was noticeably unstable 
when L310 was observed. 
Because the core is highly elliptical, the pointing fluctations induced by 
anomalous refraction could lead to underestimating the flux 
of the \c17o\ (1-0) line. 
Second, in deriving the average density in L310, 
Bacmann et al. (\cite{bacmann}) neglected the high eccentricity of the core.  
They took as size scale the dimension along its major axis, which leads 
to an underestimate in their derivation of the density.
The L310 point should be actually shifted in the direction 
of a lower depletion degree and a higher density, as indicated by the arrow 
in the graph, so that the L310 point probably lies closer to 
the rest of the data. Therefore, the depletion degree 
appears to increase with density, following a power law 
close to $f\propto n(\htwo)^{0.4-0.8}$. 

However, one has to be cautious about the ``exact'' power law 
coefficient, given the uncertainties in the H$_2$ density, due for the most
part to the uncertainty in $\kappa_{1.3\rm mm}$.
The above relation was derived assuming a constant dust opacity coefficient
and one may wonder how the exponent would be affected in the case 
$\kappa_{1.3\rm mm}$ increases with density. 
In such a case, the \htwo\ column density differs from the evaluation 
by Bacmann et al. (2000),  which, 
in turns, affects the estimation of the depletion degree too. 
We adopt a power law 
$\kappa_{1.3\rm mm} \propto n(\htwo)^{\beta}$, where $\beta\sim 0.2$ 
under the standard assumption that $\kappa_{1.3\rm mm}$ increases by a 
factor of 2 (from 0.005 to 0.01 cm$^2$g$^{-1}$) when the density increases 
from $10^5$ to $10^6$\,cm$^{-3}$ (Preibisch et al. 1993). 
A simple algebra shows that the corrected 
slope of the $f - n(\htwo)$ relation is then $\alpha(1+\beta)-\beta$ instead
of $\alpha$ ($\alpha$ is estimated, like before, supposing 
$\kappa_{1.3\rm mm}$ constant).  
The corrected slope is less steep when $\alpha \leq 1$, and it is positive  
if $\alpha$ is larger than $\beta/(1+\beta)\simeq 0.16$. The slope $\alpha$
derived previously (0.4--0.8) lies well above this value of 0.16, therefore
the corrected slope is positive (i.e. the depletion factor is found to increase
 with density, even in the case $\kappa_{1.3\rm mm}$ also increases with 
density). 
We conclude that the observed variations of the depletion factor with density 
cannot be accounted for by variations in $\kappa_{1.3\rm mm}$ only. 
Estimates of 
\htwo\ column density using the Near Infrared Color Excess method 
(cf. Bergin et al. \cite{bergin}) that are independent of 
$\kappa_{1.3\rm mm}$ should help confirm this trend. 

Assuming that a CO molecule in the gas phase has a probability equal
to one to stick onto the grain after a collision (e.g. L\'eger \cite{l83}), 
the CO
depletion is expected to vary roughly as 
$\exp\left(t/8000{\rm yr}\right)\exp(n({\rm H_2}))$, 
where $n({\rm H_2})$ is expressed in units of 
$10^5$\,cm$^{-3}$. We plotted 
in Fig.\,\ref{density} this relation for $t=2\times 10^3$ yr and 
$t=2\times 10^4$ yr (dashed lines), as well as a ``fit'' 
for $t=8\times 10^3$ yr (dash-dotted line). The observed
scatter between the data and the model may well reflect the difference in 
evolutionary stages of the pre-stellar cores studied here.  The 
dash-dotted model is consistent with the results of Caselli et al. 
(\cite{caselli}) on L1544. We note that
according to the models, the core ages are found relatively small compared
to ambipolar diffusion timescales (Ciolek \& Basu \cite{cb01}, Caselli et
al. \cite{cas02b}).
 We stress however that this simple
model does not take desorption from the grains into account, and therefore the
ages might be underestimated.

A more detailed study of the relation with 
depletion and density within the cores will be published elsewhere.

\begin{acknowledgements} 
We thank the IRAM staff in Spain for assistance in the observations and the
IRAM TAC for their award of observing time. A.B. was supported during this
work by the
\emph{Deut\-sche For\-schungs\-ge\-mein\-schaft.} P. Andr\'e is acknowledged
for his participation in the continuum observations used in this paper.
\end{acknowledgements}

\end{document}